\documentclass[conference]{IEEEtran}
\IEEEoverridecommandlockouts

\usepackage{cite}
\usepackage{tabularx}
\usepackage{amsmath,amssymb,amsfonts}
\usepackage[table,xcdraw]{xcolor}
\usepackage{algorithmic}
\usepackage{graphicx}
\usepackage{textcomp}
\usepackage{cleveref} 
\usepackage{url}
\usepackage{adjustbox}

\usepackage{amsmath}
\usepackage{algorithm}
\usepackage{algorithmic}
\usepackage{float} 
\usepackage{xcolor}
\usepackage{array}
\begin{document}
\title{


Cluster-BPI: Efficient Fine-Grain Blind Power Identification for Defending against Hardware Thermal Trojans in Multicore SoCs
\thanks{The authors would like to thank the following funding agencies: NSF grants 2219679
and 2219680.}
 }


\author{
    \IEEEauthorblockN{Mohamed R. Elshamy\IEEEauthorrefmark{1}, Mehdi Elahi\IEEEauthorrefmark{2}, Ahmad Patooghy\IEEEauthorrefmark{2}, and Abdel-Hameed A. Badawy\IEEEauthorrefmark{1}}
    \IEEEauthorblockA{\IEEEauthorrefmark{1}Klipsch School of ECE, New Mexico State University, Las Cruces, NM 88003, United States}
    \IEEEauthorblockA{\IEEEauthorrefmark{2}Computer Systems Technology, North Carolina A\&T State University, Greensboro, NC, United States}
    \IEEEauthorblockA{\IEEEauthorrefmark{1}\{elshamy, badawy\}@nmsu.edu, \IEEEauthorrefmark{2}melahi@aggies.ncat.edu, apatooghy@ncat.edu}
}

\maketitle

\begin{abstract}
Modern multicore System-on-Chips (SoCs) include hardware monitoring mechanisms to measure total power consumption, but these aggregate measurements are insufficient for fine-grained thermal and power management. This paper introduces an improved Clustering Blind Power Identification (ICBPI), an approach to improve the sensitivity and robustness of the Blind Power Identification (BPI) approach, which identifies the power consumption of different cores and the thermal model of an SoC using only thermal sensor measurements and the total power consumption. The proposed approach enhances BPI's initialization step (specifically the non-negative matrix factorization, which is crucial for BPI accuracy) by incorporating density-based spatial clustering of noise applications (DBSCAN). This maximizes the physical relationship between the temperature and power consumption, ensuring more accurate power estimates. Our simulations demonstrate two tasks to validate the proposed approach. The first evaluates the power accuracy per core on four different multicores, including a heterogeneous processor, showing that ICBPI significantly improves accuracy without overheads. For example, in a four-core SoC, error rates are reduced by 77.56\% compared to vanilla BPI and by 68.44\% compared to the state-of-the-art approach called BPISS. The second task focuses on enhancing the precision and robustness of the detection and localization of malicious thermal sensor attacks in the heterogeneous processor, demonstrating that ICBPI can enhance the security of multicore SoCs.
\end{abstract}


\section{Introduction}
As the progress of Moore's law slows down, designers have turned to alternative design methodologies. This shift has led to the development of heterogeneous multicore architectures and the integration of specialized hardware units into a single chip, known as System-on-Chips (SoCs) ~\cite{https://doi.org/10.1002/cpe.1904}. However, these advances have brought about challenges related to thermal management, power consumption, and energy efficiency~\cite{b1}, e.g., increased power density and limited cooling options in SoCs have resulted in significant performance bottlenecks~\cite{RANGARAJAN2023185}. Consequently, dynamic Thermal Management (DTM) techniques have become essential in all mobile SoCs to reduce and control high operating temperatures~\cite{7544290}. 

Accurate estimation of the power consumption of each core in a multicore SoC is critical to effective thermal and power management~\cite{b4,10.1007/978-3-030-60939-9_13}. However, modern processors lack fine-grained power sensors. For example, the running average power limit (RAPL) interface allows applications to measure power consumption~\cite{5599016}. However, these measurements are generally coarse-grained, providing only the overall power consumption of all cores, uncore units, and total package power. Fine-grained power estimation improves the efficiency of dynamic voltage and frequency scaling (DVFS) and thermal throttling, maintaining safe temperature limits and enhancing processor reliability, lifespan, and cooling costs~\cite{b4,b5}.

Blind Power Identification (BPI) is a recent approach~\cite{b4} that can simultaneously estimate the fine-grained power consumption of each core in multicore SoCs and identify the thermal model of the chip using only thermal sensor measurements and total power consumption data. 
The effectiveness of the BPI algorithm is highly sensitive to dataset dimensions, values, and outliers due to its dependence on the non-negative matrix factorization (NMF) technique, which is particularly susceptible to initialization~\cite{b13}. This sensitivity poses challenges in environments where precise thermal management is crucial, such as SoCs used in mobile applications. Traditional NMF initialization approaches are often too sensitive to dataset variations, leading to inconsistent results and poor accuracy in estimating fine-grained power~\cite{b6}. In addition, NMF sensitivity to thermal outliers distorts factorization, resulting in suboptimal matrix approximations and degraded feature quality~\cite{hafshejani2023initialization}.

Advancements in digital systems have led many IC companies to adopt fabless production, outsourcing various stages to reduce costs and time to market~\cite{4702884}. However, this increases the risk of vulnerabilities, such as hardware Trojans (HTs) that compromise thermal sensors in SoCs, which are critical for DTM systems~\cite{6856140}. Techniques like BPI for fine-grained power estimation can help mitigate these threats and are essential for the future of SoC design.

This paper proposes Improved Clustering BPI (ICBPI). This approach automates NMF initialization using DBSCAN to identify dense regions and initialize the NMF approach, helping to avoid local minima and enhance robustness. Furthermore, ICBPI addresses scenarios where a thermal sensor is maliciously or benignly failing, which can result in performance degradation or reduced chip lifespan due to excessive frequency throttling or accelerated aging. Our research demonstrates that ICBPI, the proposed approach, significantly improves power estimation accuracy across multicore SoCs, including homogeneous and heterogeneous architectures. This automated fine-grained power estimation ensures accuracy regardless of variation in the temperature data, optimizes chip performance, and improves its security against thermal attacks. ICBPI's contributions pave the way for future power and thermal management advancements and security measures for SoCs.

The remainder of this paper is structured as follows. Sections II and III cover related work and background information. Section IV presents the proposed ICBPI approach. Section V discusses the results and compares our approach with existing techniques. Section VI concludes the paper.

\section{Related Work}
Numerous studies have explored thermal and power modeling to detect and mitigate hardware Trojan attacks on multicore SoCs~\cite{6629264,9806243}.
This paper focuses on blind power identification~\cite{b4,b5,b6}. 

Reda~\textit{et al.}~\cite{b4} propose the first version of the BPI approach, which used the fast Independent Component Analysis (ICA) algorithm~\cite{761722} to initialize NMF. This approach provides an initial estimate of the factors that enhance convergence and performance by ensuring better separation of the underlying sources. Also, Reda~\textit{et al.}~\cite{b5} proposed an improved BPI version with a new NMF initialization using the identity matrix for the thermal resistance matrix. This method initializes the power matrix, $P$, by equally distributing the total power among the cores, respecting the self-coupling of the thermal resistance matrix, and achieves a higher precision of fine-grained power estimation compared to the previous ICA approach.

Said~\textit{et al.}~\cite{b6} proposed BPISS, an enhancement that initializes the thermal resistance matrix by setting non-diagonal elements to the average steady-state temperatures of stressed cores and diagonal elements to cores with similar thermal characteristics. The power matrix is initialized by dividing the total power based on each core's temperature ratio. This approach better reflects the system's physical characteristics and reduces error compared to identity matrix initialization. However, previous methods do not guarantee the avoidance of local minima or account for outliers in the dataset, leading to suboptimal solutions. To address these issues, this paper proposes the ICBPI approach.

Recent advances have focused on detecting and mitigating hardware Trojans and malicious attacks on thermal sensors. Zhang \textit{et al.}~\cite{10.1145/2660267.2660289} developed DETRUST, a technique that uses stealthy, implicitly triggered hardware Trojans to bypass trust verification mechanisms. Sebt \textit{et al.}~\cite{https://doi.org/10.1049/iet-cdt.2018.5108} identified vulnerabilities at the gate level, exposing circuit enclaves to hardware Trojan insertions. In mobile SoCs, Abdelrehim \textit{et al.}~\cite{9806243} introduced a blind identification countermeasure (BIC) to detect and mitigate malicious thermal sensor attacks with high accuracy and minimal performance impact. Since BIC relies on BPI for accurate SoC matrix estimation, this paper proposes ICBPI to improve these estimations and strengthen BIC's effectiveness against thermal sensor attacks.

\section{Background}
This section provides a background on the BPI and BIC approaches, which will be improved in this paper using the proposed ICBPI approach.

\subsection{BPI Approach}
The fine-grained power estimation is based mainly on the BPI approach, which, in turn, is based on the standard methodology for thermal and power modeling of multicores, known as the state-space model~\cite{5763141} as noted in Eq. \ref{Eq:1}
\begin{equation}
    T_r(k) = AT_r(k-1) + BP(k).
    \label{Eq:1}
\end{equation}
where, \( T_r(k) \) and \( P(k) \) represent matrices denoting the temperature and power levels of SoC units at time \( k \), respectively, the matrices \( A \) and \( B \) encapsulate the physical relationship between power and temperature. While matrix \( A \) is the thermal conductance matrix that illustrates the natural response of the system in the absence of power input, matrix \( B \) describes the forced response of the system as a function of thermal capacitance and conductance. 

\begin{figure}[tb]
\centering
\includegraphics[width=0.85\linewidth]{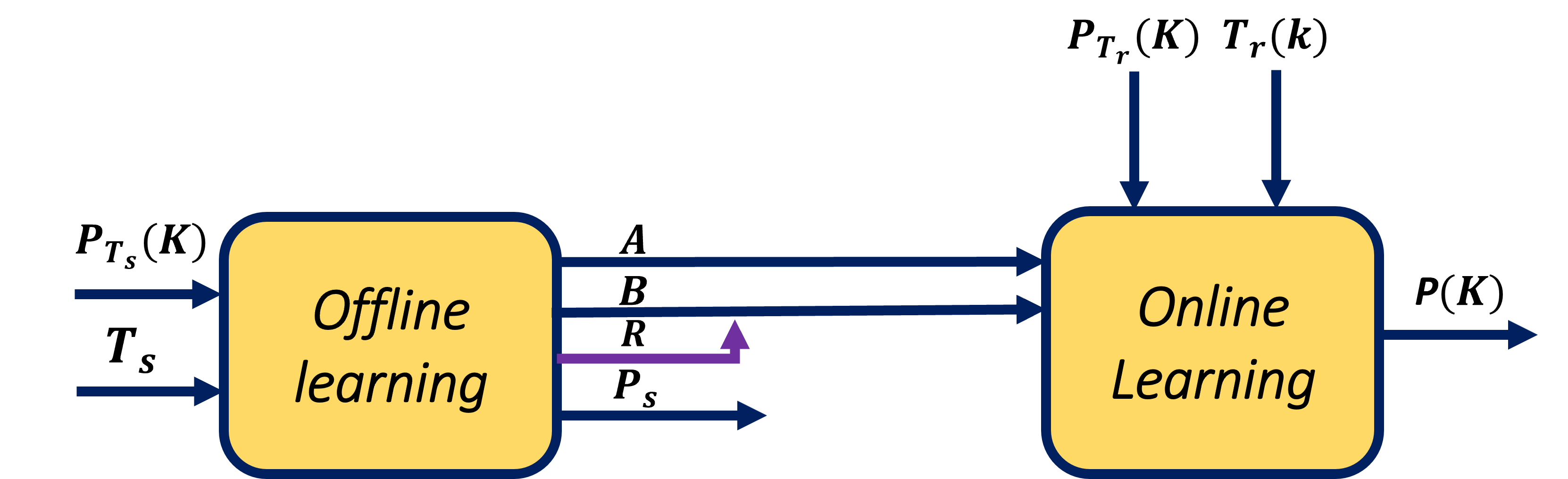}
\vspace{-0.3cm}
\caption{Interactions and data-flow of the Blind Power Identification Algorithm}
\label{fig:BPI_Phases}
\end{figure}

The BPI approach operates in two phases, as shown in Fig.~\ref{fig:BPI_Phases}. The first phase is offline learning, where the matrices \(A\), \(B\), and \(R\) are estimated using steady-state measurements, $T_s$, and the total power, $P_{T_{s}}(k)$. The second phase is online learning, which dynamically estimates core power consumption during runtime using $T_r(k)$ and $P_{T_{r}}(k)$, enabling real-time adjustments~\cite{b5}. To compute the matrix \(B\) in the first phase, it is necessary to determine the matrix $\mathbf{R} \in \mathbb{R}^{N \times N}$ (the thermal resistance matrix) in the steady state (SS) scenario~\cite{b5}, where \(T_r(k) = T_r(k-1)\). Therefore, the following equation can be derived from Eq.~\ref{Eq:1}:

\vspace{-12pt} 
\begin{align}
T_s & \approx R P_s. \label{eq:5}
\end{align}

Although $\mathbf{T}_S$ is known, only the total power, $\mathbf{P}_S$, is accessible due to the high cost and limited number of power sensors. BPI estimates both $\mathbf{R}$ and $\mathbf{P}_S$ using NMF. The proposed ICBPI approach improves the initialization of the NMF, thus improving the power estimation accuracy per unit.

\subsection{BIC Approach}

The BIC approach fundamentally depends on the BPI approach, which uses the traditional identity matrix initialization for NMF~\cite{b5}. It is also influenced by the matrix \(R\), determined by the layout geometry, materials, and the properties of the cooling system~\cite{9806243}. Mobile SoCs typically use passive phase change material cooling~\cite{KURHADE20213171}, ensuring that it remains consistent once \(R\) is determined. This principle forms the foundation of BIC, which consists of four phases. In \textbf{Phase I}, \( R_{\text{Golden}} \) is established by offline tests with calibrated sensors. During \textbf{Phase II}, \( R_{\text{Runtime}} \) is recomputed at runtime and compared to \( R_{\text{Golden}} \), with deviations beyond a specified tolerance \( \xi \) being checked. In \textbf{Phase III}, the malicious sensor is identified by comparing the submatrices of \( R_{\text{Runtime}} \) and \( R_{\text{Golden}} \). Finally, in \textbf{Phase IV}, the temperature of the compromised sensor is estimated to maintain effective thermal management.

An accurate estimation of the matrix \( R \) is essential for effective attack detection in BIC and is highly dependent on the initialization of the NMF. The proposed ICBPI approach improves this process, enhancing detection and identification. This paper follows the same threat model as the original study~\cite{9806243}.

\section{The Proposed Approach}
\label{Sec:Proposed-Init}
This section describes the proposed ICBPI approach to accurately estimate thermal modeling matrices and per-unit power in multicore processors. The natural response matrix \( A \) is estimated by setting \( P(k) = 0 \) in Eq.~\ref{Eq:1}, which leads to equation \( T_r(k) = A T_r(k-1) \). By collecting $K$ thermal traces to construct matrices $\mathbf{T1} = [T_r(1) \cdots T_r(K-1)]$ and $\mathbf{T2} = [T_r(2) \cdots T_r(K)]$, the quadratic programming problem is solved as $\min \| \mathbf{T2} - \mathbf{A} \mathbf{T1} \|_F^2$, subject to $\mathbf{A} \geq 0$. Here, all entries in $\mathbf{A}$ are non-negative. The \( R \) and \( B \) matrices are then estimated from \( T_s \) using NMF. The ICBPI approach is applied at this step to obtain the most accurate \( R \). Finally, the runtime power consumption is estimated by solving another quadratic programming problem.

\begin{equation}
\begin{aligned}
\min &\quad ||BP(k) - (T_r(k) - AT_r(k-1))||_2^2 \\
\text{subject to} &\quad P(k) \geq 0 \\
&\quad ||P(k)||_1 = \text{total measured power}.
\end{aligned}
\end{equation}

The proposed ICBPI approach utilizes the DBSCAN algorithm to initialize the input matrix of the NMF \( R \), improving the precision of the per-core power estimation. The thermal resistance matrix \( R \) represents the impact of heat generated in one core on the temperature of other cores. For example, on a chip with a \(2 \times 2\) core, the \(4 \times 4\) matrix \( R \) includes elements such as \(r_{12}\), which indicates the thermal influence of core one on core two. The proposed ICBPI approach uses DBSCAN clustering to identify hotspot areas in the SoC dataset, pinpointing regions of the highest power dissipation. The centroids of these areas are used to initialize the rows of the \( R \) matrix, improving its representation of thermal impact. In Eq.~\ref{eq:5}, \( R \) is directly related to \( T_s \), which means that changes in \( P \) affect \( T_s \), and in a steady state, variations in \( R \) adjust \( P_s \). Initializing \( R \) with the optimal points in the \( T \) dataset leads to a more accurate NMF convergence. To clarify the initialization of the matrix \( R \) with centroids of the hotspot area, consider an example of a 4-core SoC and activate the fourth core that allows the proposed ICBPI to detect the hotspot areas on the chip using DBSCAN clustering. Assuming there is one hotspot area, the centroid of this area is denoted as \([r_{11}, r_{12}, r_{13}, r_{14}]\) because it is a four-core processor. 

\begin{figure}[tb]
\centering
\includegraphics[scale=0.36, trim=150 0 0 100, clip]{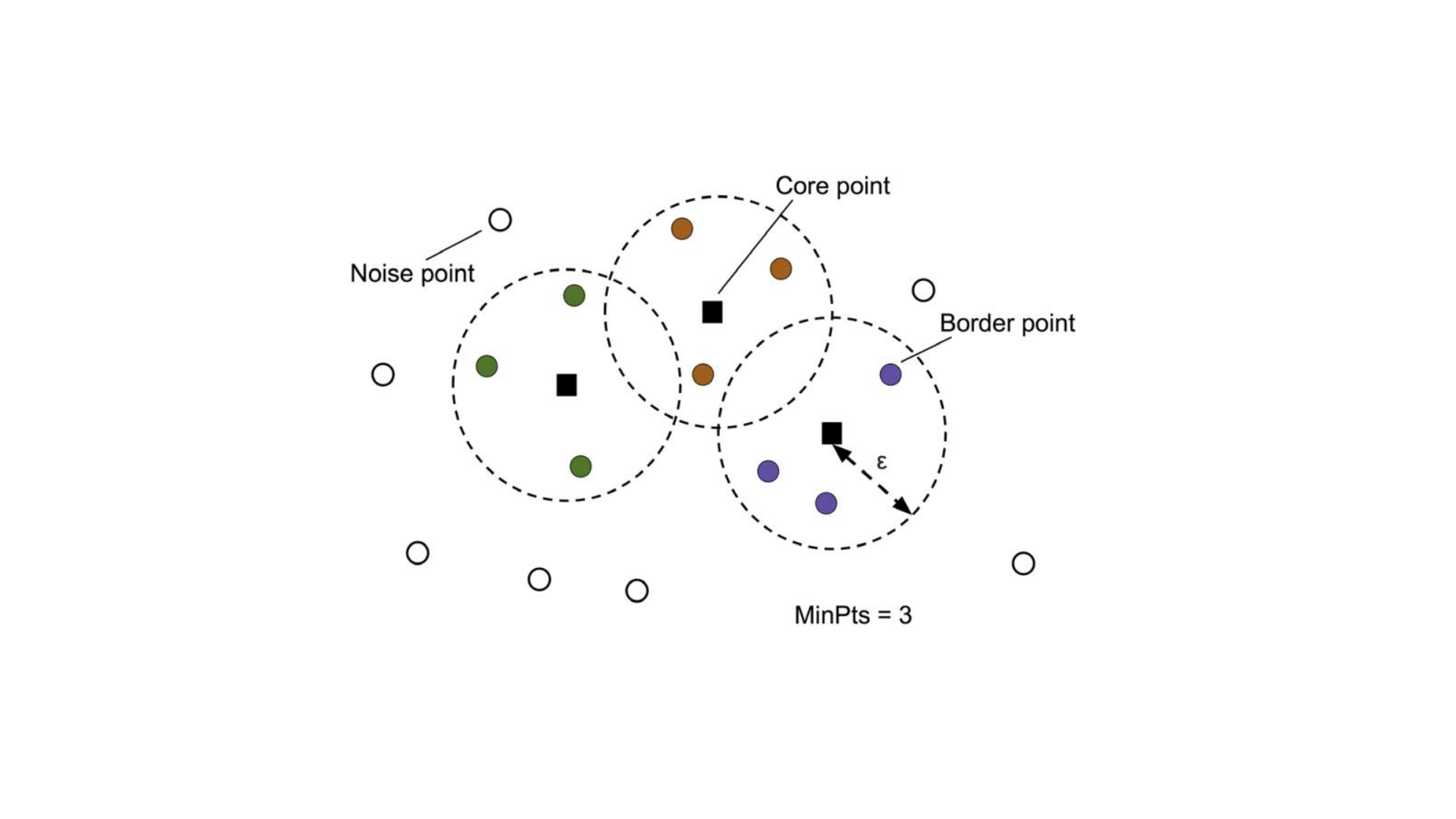}
\vspace{-2.0cm}
\caption{Illustration of the DBSCAN Clustering 
Algorithm~\cite{geron2022machine}}
\label{fig:DBSCAN_Points}
\end{figure}

DBSCAN is highly efficient at identifying clusters of various shapes and sizes within a dataset, making it particularly effective for detecting active cores compared to other clustering techniques. As shown in
Fig.~\ref{fig:DBSCAN_Points}, DBSCAN typically uses the Euclidean distance to measure the distance between points~\cite{10.5555/3001460.3001507,b7,10392256},  
$d(p, q) = \sqrt{\sum_{i=1}^{n} (p_i - q_i)^2}$.   
Here, \( p \) and \( q \) are two points in a \( n \) dimensional space. A core point is a point \( p \) that has at least \textit{MinPts} points distanced \( \epsilon \) from it (including \( p \)), that is, \( |N_\epsilon(p)| \geq \text{MinPts} \) and \( N_\epsilon(p) \) is the set of points within \( \epsilon \) distance from \( p \). A point \( q \) is directly density accessible from a point \( p \) if \( p \) is a core point and \( q \) is within the distance \( \epsilon \) from \( p \), that is, \( q \in N_\epsilon(p) \). A point \( q \) is density-reachable from \( p \) if there is a chain of points \( p_1, p_2, \ldots, p_n \), with \( p_1 = p \) and \( p_n = q \), where each point is directly density-reachable from the previous one. A core point has at least MinPts neighbors within a radius \(\epsilon\), while a border point has fewer neighbors but is near a core point. Noise points are neither core nor border points and are considered outliers. DBSCAN clusters data based on the density of points within \(\epsilon\), starting from a core point and expanding recursively to all density reachable points. It handles noise well and can identify clusters with arbitrary shapes, but its precision depends on selecting the appropriate \(\epsilon\) and MinPts.

\begin{figure}[tb]
\centering
\includegraphics[scale=0.38, trim=180 50 0 60, clip]{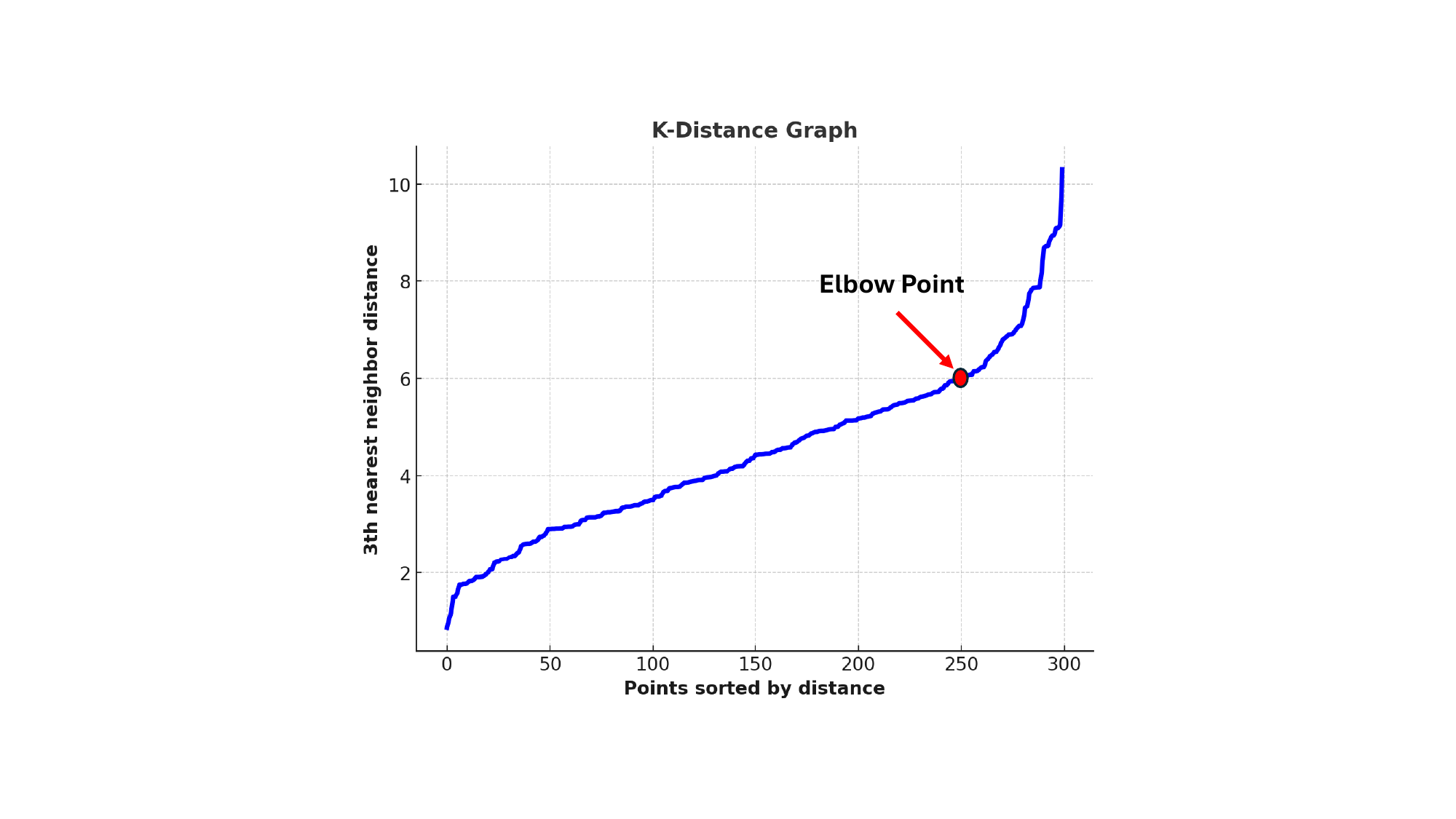}
\vspace{-10mm}
\caption{K-distance graph to determine the value of \(\epsilon\)}
\label{fig:K-Distance}
\end{figure}

To estimate the optimal \(\epsilon\) for DBSCAN, the k distance graph is used, plotting the distance to the k-th nearest neighbor in descending order. The "elbow point" on the graph, where the slope changes significantly, indicates the optimal \(\epsilon\), as shown in Fig.~\ref{fig:K-Distance}. A general rule of thumb is to set MinPts at least $D+1$, where $D$ is the number of dimensions in the dataset, ensuring sufficient data for cluster formation. The detailed approach is described in~\Cref{alg:DBSCAN_BPI}.

\begin{algorithm}[tb]
\caption{ICBPI}\label{alg:dbscan_bpi}
\begin{algorithmic}[1]
    \STATE \textbf{
    ICBPI Offline Phase:}
    \STATE \textbf{Input:} Steady-state thermal traces $T_s$, and corresponding total power at each time interval $P_{T_{s}}(k)$.
    \STATE Assume $\mathbf{T1} = [T_r(1) \cdots T_r(K-1)]$ and $\mathbf{T2} = [T_r(2) \cdots T_r(K)]$. 
    \STATE Calculate The quadratic programming problem: $\min \| \mathbf{T2} - \mathbf{A} \mathbf{T1} \|_F^2$ subject to $\mathbf{A} \geq 0$.
    \STATE Determine the optiumum value of \(\epsilon\) using k-distance graph.
    \STATE Adjust MinPts to be at least D + 1, where D is the number of dimensions (features) in your dataset.
    \STATE Apply the DBSCAN to $T_s$ to identify clusters based on the dense regions in $T_s$ and remove outliers.
    \STATE Initialize $R$ with the centroids of the DBSCAN clusters.
    \STATE Initialize $P_s$ with $T_s$.
    \STATE Apply the NMF with the proposed initialization.
    \STATE \textbf{Output:} System matrices $A$, $B$, and $R$
    \STATE \textbf{BPI Online Phase}  
\end{algorithmic}
\label{alg:DBSCAN_BPI}
\end{algorithm}


\section{Experiments and Results }
\subsection{Experimental setup}
\begin{figure}[tb]
\centering
\includegraphics[scale=0.32, trim=355 480 0 240, clip]{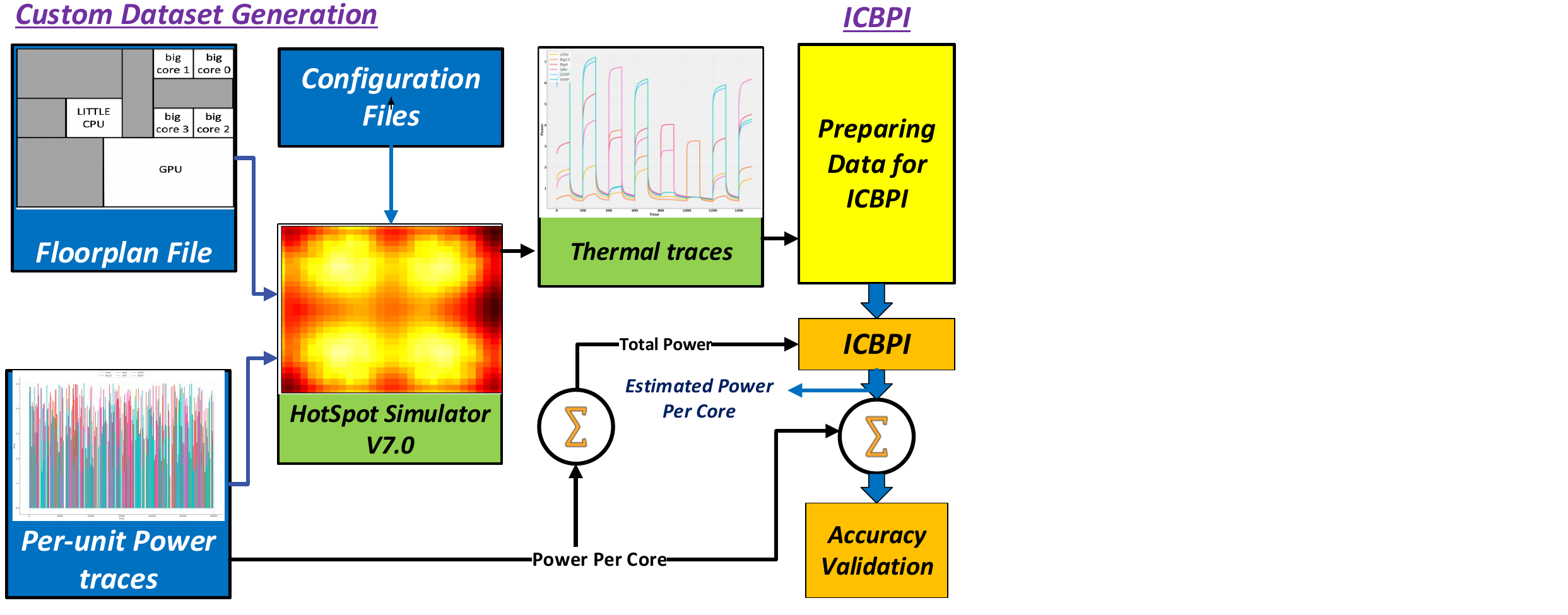}
\vspace{-6mm}
\caption{ICBPI Custom Dataset Generation and Validation Workflow.}
\label{fig:ICBPI_Algorithm}
\end{figure}

The effectiveness of the ICBPI approach is validated using the HotSpot thermal simulator v7~\cite{HotSpot} in four different floorplans, as detailed in Table~\ref{table:floorplans}. Custom datasets were generated for each floorplan, and the power trace, along with the floorplan structure and the HotSpot configuration files, were fed into the simulator to obtain the thermal trace as shown in Fig.~\ref{fig:ICBPI_Algorithm}. The ambient temperature was set at $298.15K$, with a sampling interval of 100 milliseconds, selected to match the steady-state temperature behavior of the processor.

\begin{table}[tb]
\centering
\vspace{-3mm}
\caption{Floorplans with Unit Counts and Power Budgets (Area: 1 cm\(^2\))}
\vspace{-2mm}
\label{table:floorplans}
\begin{tabular}{|>{\centering\arraybackslash}p{3cm}|>{\centering\arraybackslash}p{2cm}|>{\centering\arraybackslash}p{2cm}|}
\hline
\textbf{Floorplans} & \textbf{Units} & \textbf{Power Budget} \\ \hline
$2 \times 2$ mesh & 4 & 80W \\ \hline
$2 \times 4$ mesh & 8 & 80W \\ \hline
$4 \times 4$ mesh & 16 & 80W \\ \hline
big.LITTLE+GPU~\cite{b14} & 6 & 15W \\ \hline
\end{tabular}
\end{table}

The proposed approach will be evaluated using two tasks. \textbf{The first task} applies the generated dataset to both the proposed approach and the state-of-the-art methods from the literature. The accuracy is assessed by calculating the average absolute error percentage between the estimated per unit of power and the actual per unit power traces provided to the HotSpot simulator, as shown in Eq.~\ref{Eq:5}.

\vspace{-5mm}
\begin{equation}
\text{av. abs. error (\%)} = \frac{1}{N} \sum_{n=1}^{N} \frac{\left| \text{estimated power} - \text{actual power} \right|}{\text{actual power}}
\label{Eq:5}
\end{equation}

where N is the number of cores. 


\textbf{The second task} uses the ICBPI approach to enhance the BIC algorithm for detecting and localizing thermal attacks. The setup involves a custom dataset for the heterogeneous floorplan, where power traces and layout information are fed into the HotSpot 7.0 simulator to generate thermal traces. BIC is tested by considering one core's thermal sensor as malicious. The malicious sensor is modeled by altering the HotSpot temperature with \( \Delta t_{\text{error}} \). The value of \( \Delta t_{\text{error}} \) is iteratively changed within the ranges of \( (-15:-1) \) and \( (1:15) \), with a 1°C step. Additionally, \( \xi \) is varied for the heterogeneous layout from \( (0.01:0.1) \), with a 0.01 step.

\begin{figure}[tb]
    \centering
    \includegraphics[scale=0.3, trim={255 0 0 220}, clip]{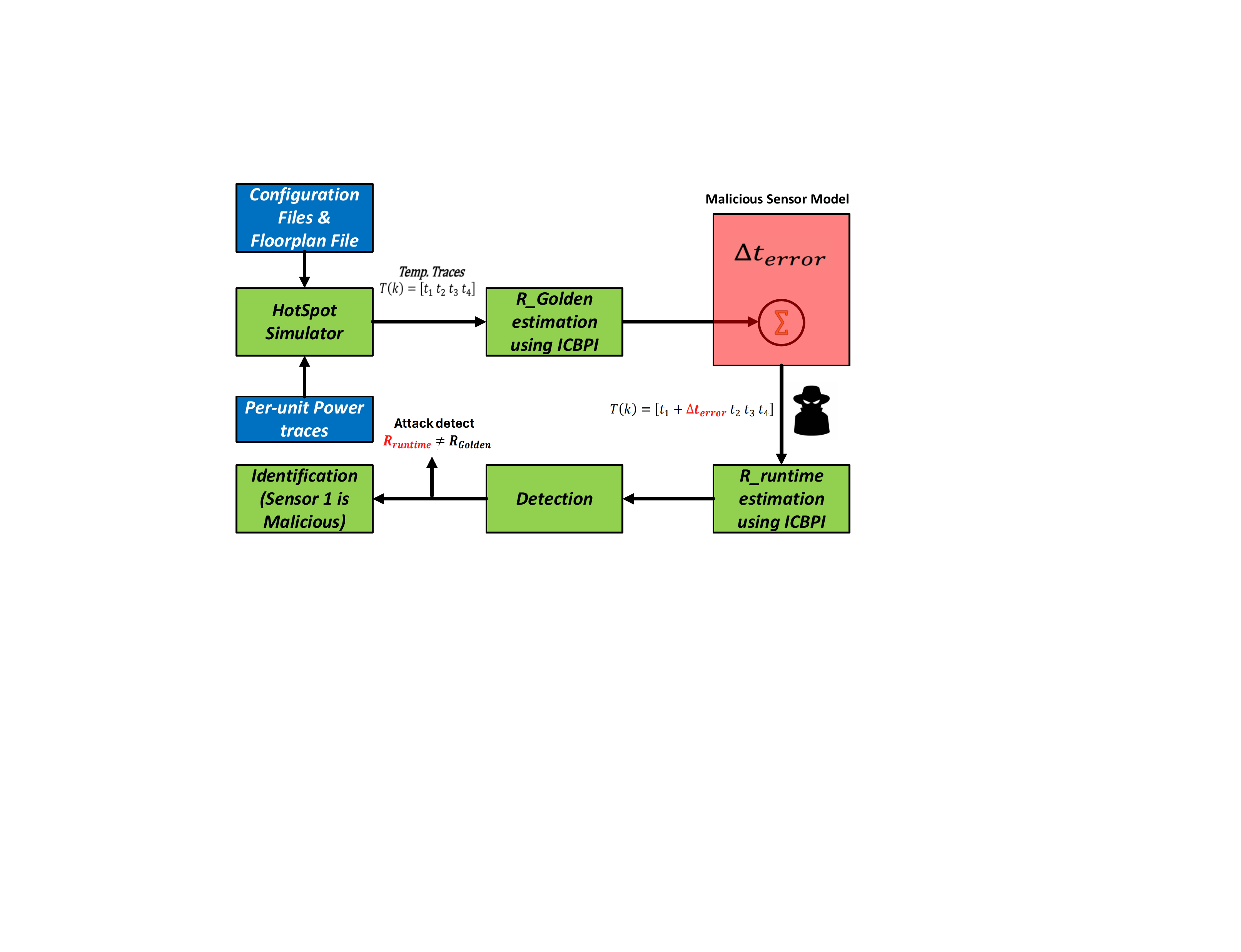}
    \vspace{-5.7cm}
    \caption{The simulation models a 4-core processor where thermal sensor `1` is made malicious by introducing an error to its reading, \( t_1 \).}
    \label{fig:BIC1}
\end{figure}

For each tuple $(\xi, \Delta t)$, the simulator tracks the number of failures, where failure is defined as the BIC that does not detect the attack or identify the attacker. The maximum number of failures for any tuple is the number of cores, $N$. An instance where Core 1 harbors the malicious sensor is illustrated in Fig.~\ref{fig:BIC1}. The generated dataset will be applied to both the proposed ICBPI approach and the state-of-the-art BIC, and the results will be compared.


\subsection{Task 1: Evaluation of the proposed approach using the generated dataset}The comparison results, summarized in Table~\ref{table:power_estimation_Comparison} and Table~\ref{table: runtime}, show the improvements of the ICBPI approach over the state-of-the-art methods. ICBPI was validated on a heterogeneous architecture (big.LITTLE+GPU), demonstrating its robustness in both accuracy and computation time when transitioning from homogeneous to heterogeneous systems. It achieved higher accuracy, particularly with larger benchmarks, indicating scalability. Although traditional approaches have slightly lower runtime, they suffer from higher estimation errors, whereas BPISS shows the highest runtime. ICBPI balances runtime and accuracy, significantly reducing estimation errors.

\begin{table}[tb]
\caption{The power estimation error of the ICBPI compared to BPI and BPISS using four floorplan benchmarks.}

\centering
\centering
\begin{tabularx}{\columnwidth}{|p{1.5in}|X|X|X|}
\hline
Benchmarks & BPI~\cite{b5} & BPISS~\cite{b6} & ICBPI \\
\hline
2x2 mesh & 4.50 & 3.20 & 1.01 \\
\hline
2x4 mesh & 11.10 & 6.98 & 3.24 \\
\hline
4x4 mesh & 3.80 & 2.80 & 0.51 \\
\hline
big.LITTLE+GPU& 5.04 & 3.90 & 0.71 \\
\hline
\end{tabularx}
\label{table:power_estimation_Comparison}
\end{table}

\begin{table}[tb]
\caption{The runtime in seconds of ICBPI compared to BPI and BPISS across four different floorplans.}
\centering
\begin{tabularx}{\columnwidth}{|p{1.5in}|X|X|X|}
\hline
Benchmarks & BPI~\cite{b5} & BPISS~\cite{b6} & ICBPI \\
\hline
2x2 mesh & 14.19 & 16.75 & 13.24 \\
\hline
2x4 mesh & 20.27 & 23.50 & 19.19 \\
\hline
4x4 mesh & 27.19 & 30.23 & 27.04 \\
\hline
big.LITTLE+GPU& 27.10 & 29.10 & 27.87 \\
\hline
\end{tabularx}
\label{table:runtime}
\end{table}

\begin{figure}
    \centering
    \includegraphics[scale=0.62, trim={0 0 0 0}, clip]{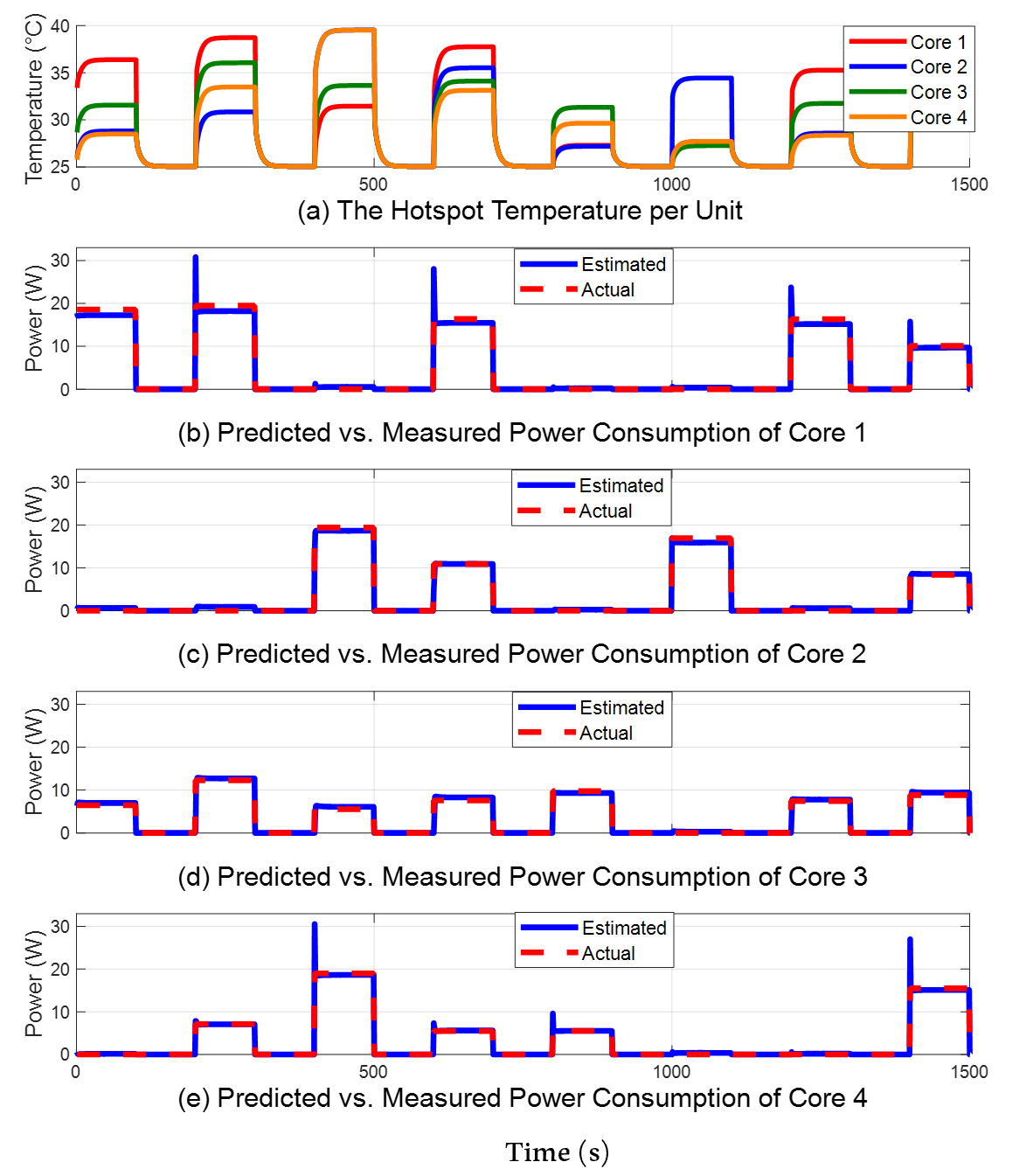}
    \vspace{-0.2cm}
    \caption{Validation of the Proposed Approach ICBPI }
    \label{fig:4_Core}
\end{figure}

To verify the power estimation capabilities of the ICBPI algorithm, a comparative analysis between estimated and actual power consumption for each core in the 2x2 mesh floorplan is presented. Fig.~\ref{fig:4_Core} shows the thermal measurements over time in sub-figure (a), highlighting fluctuations due to the varying core stress levels. The remaining subfigures compare the power estimates of ICBPI with the actual power input, demonstrating a close match. This strong correlation validates ICBPI's effectiveness in providing accurate power estimates, which is essential for optimizing power management and improving processor efficiency.

\subsection{Task 2: Enhancement of Thermal Attack Detection and Localization}

\begin{figure}
    \centering
    \includegraphics[scale=0.45, trim={ 100 150 0 180}, clip]{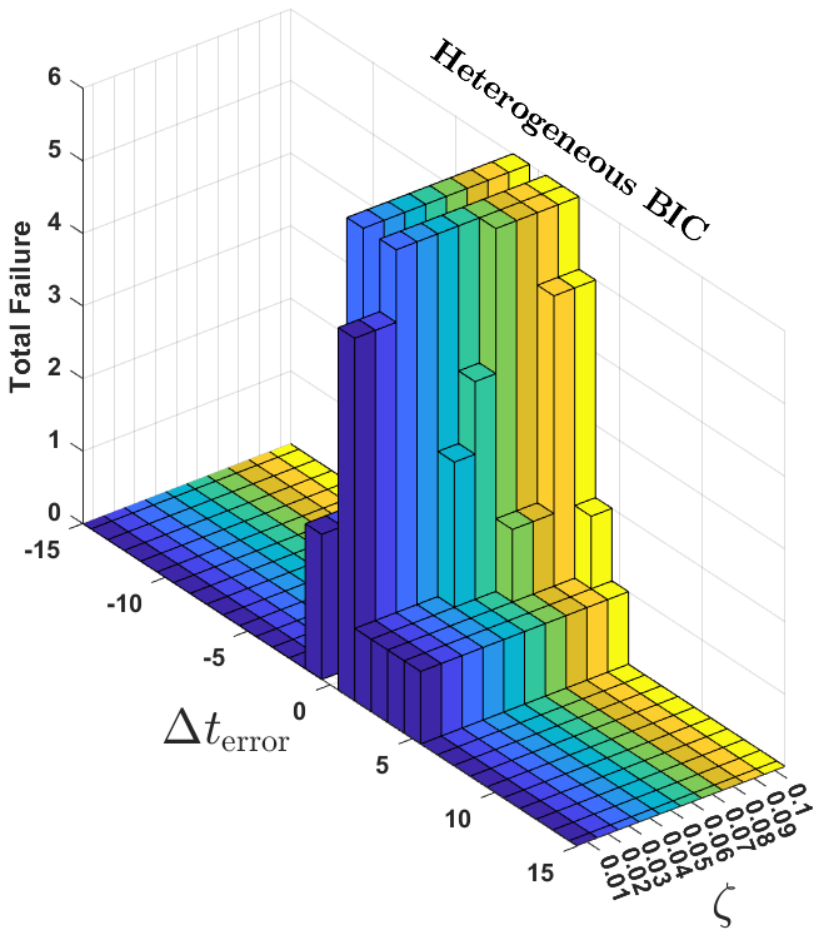}
    \caption{Relationship between failure counts and $(\Delta t_{error}, \xi)$ using standard BIC approach}
    \label{fig:BIC}
\end{figure}

\begin{figure}
    \centering
    \includegraphics[scale=0.5,  trim={100 150 0 200}, clip]
    {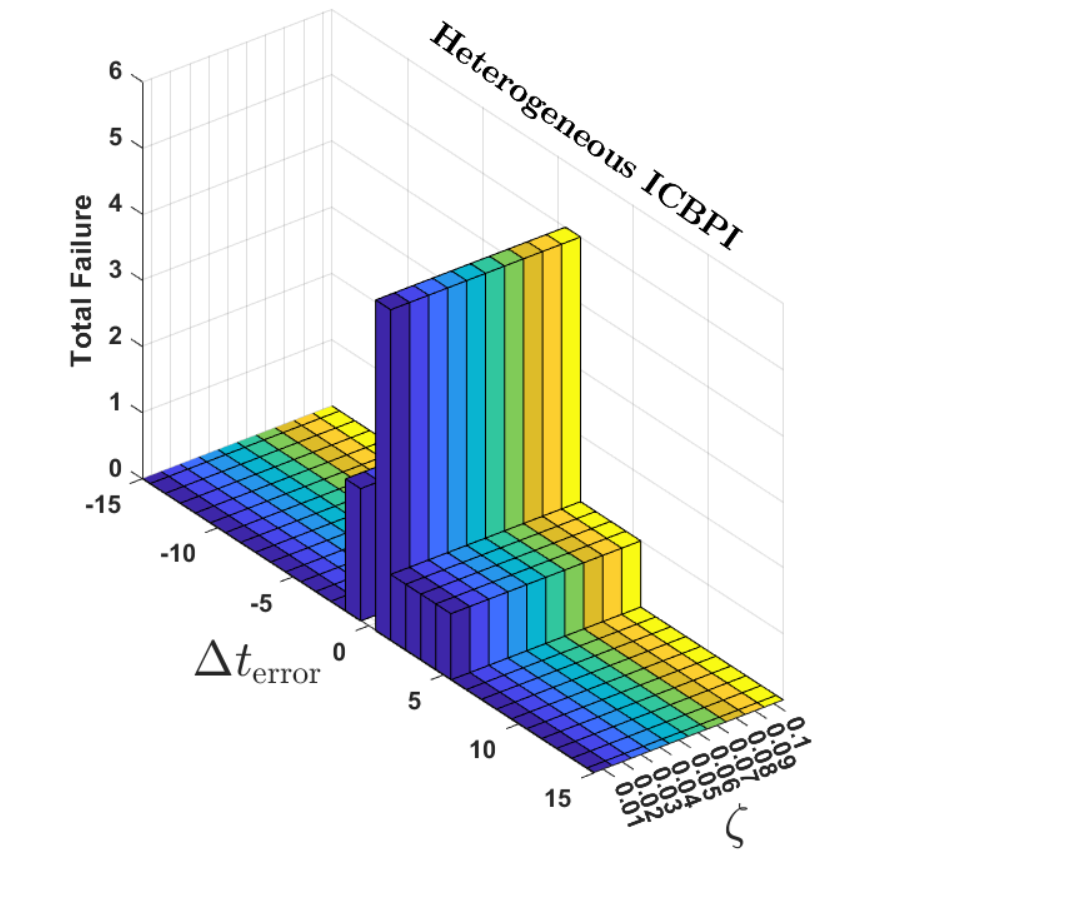}
    \caption{Relationship between failure counts and $(\Delta t_{error}, \xi)$ using the proposed ICBPI approach.}
    \label{fig:ICBPI}
\end{figure}

Fig.~\ref{fig:BIC} shows the results of the standard BIC approach, while Fig.~\ref{fig:ICBPI} presents the results after integrating ICBPI with BIC. The comparison highlights the significant performance improvement with ICBPI, which maintains high accuracy in detecting sensor attacks for small $\xi$ and reduces failures in various scenarios. This demonstrates the robustness and reliability of the ICBPI approach, marking a notable improvement over the conventional BIC technique.

A comparison between standard BIC and BIC with ICBPI, as shown in Table~\ref{tab:comparison}, reveals significant improvements in detection and identification failure rates. The standard BIC shows detection failures, reaching $27.27\%$ for $\Delta t_{\text{error}}$ values of $-1$, $1$, $2$ and $3$. In contrast, BIC with ICBPI consistently has $0\%$ detection failures for all $\Delta t_{\text{error}}$ values, with only some identification failures, such as $16.18\%$ at $\Delta t_{\text{error}} = -1$ and $30.4\%$ at $\Delta t_{\text{error}} = 1$. This demonstrates ICBPI's effectiveness in significantly reducing detection and identification failures compared to standard BIC.

\begin{table}
\caption{Comparison of Detection and Identification Failure Rates}
\centering
\scriptsize 
\setlength{\tabcolsep}{2pt} 
\begin{tabular}{|c|c|c|c|c|}
\hline
$\Delta_{t_{error}}$ & \multicolumn{2}{c|}{Heterogeneous - BIC} & \multicolumn{2}{c|}{Heterogeneous - ICBPI} \\ \cline{2-5} 
                 & Detection Failure & Ident. Failure & Detection Failure & Ident. Failure \\ \hline
-6:-15           & 0\%               & 0\%               & 0\%               & 0\%               \\ \hline
-5               & 0\%               & 0\%               & 0\%               & 0\%               \\ \hline
-4               & 0\%               & 0\%               & 0\%               & 0\%               \\ \hline
-3               & \cellcolor[gray]{0.8} 3.72\%  & 0\%               & 0\%               & 0\%               \\ \hline
-2               & \cellcolor[gray]{0.8} 13.04\% & 0\%               & 0\%               & 0\%               \\ \hline
-1               & \cellcolor[gray]{0.8} 25.467\% & \cellcolor[gray]{0.8} 20\% & 0\%               & \cellcolor[gray]{0.8} 16.18\% \\ \hline
1                & \cellcolor[gray]{0.8} 27.27\%  & \cellcolor[gray]{0.8} 27.25\% & 0\%               & \cellcolor[gray]{0.8} 30.4\% \\ \hline
2                & \cellcolor[gray]{0.8} 19.25\%  & \cellcolor[gray]{0.8} 7.27\% & 0\%               & \cellcolor[gray]{0.8} 9.09\% \\ \hline
3                & \cellcolor[gray]{0.8} 8.69\%   & \cellcolor[gray]{0.8} 10.9\% & 0\%               & \cellcolor[gray]{0.8} 9.09\% \\ \hline
4                & \cellcolor[gray]{0.8} 1.24\%   & \cellcolor[gray]{0.8} 16.36\% & 0\%               & \cellcolor[gray]{0.8} 9.09\% \\ \hline
5                & 0\%               & \cellcolor[gray]{0.8} 18.18\% & 0\%               & \cellcolor[gray]{0.8} 9.09\% \\ \hline
6:15             & 0\%               & 0\%               & 0\%               & 0\%               \\ \hline

\end{tabular}
\label{tab:comparison}
\end{table}

%

\section{Conclusion} 
The proposed ICBPI approach significantly improves the accuracy and robustness of the Blind Power Identification (BPI) approach by utilizing DBSCAN for optimal initialization of NMF, addressing its sensitivity to dataset variations and outliers. Our extensive simulations have demonstrated that ICBPI achieves higher accuracy in fine-grained power estimation across various multicore SoC configurations, including both homogeneous and heterogeneous architectures. In particular, error rates in a 4-core processor were reduced by $77.56\%$ compared to traditional BPI approaches and by $68.44\%$ compared to the state-of-the-art BPISS approach. Using the ICBPI approach within the BIC security approach, ICBPI demonstrated superior performance in detecting and localizing malicious thermal sensor attacks. This method maintains high accuracy and reduces failure rates in various scenarios, outperforming the original BIC approach. However, the main limitation lies in detecting simultaneous attacks in the SoC.


\section*{Acknowledgments}
The authors thank Dr. Moustafa Abdelrahim of Qualcomm Inc.\ for providing codes from his BIC paper (ISQED'2022)~\cite{9806243}. Also, the authors would like to thank the anonymous reviewers. 
This work has been partially funded by NSF grants 2219679 and 2219680.

\bibliography{main} 

\end{document}